\documentclass[prd,amsmath,amssymb,showpacs,superscriptaddress,nofootinbib,twocolumn]{revtex4-1}
\usepackage{graphicx}
\usepackage{epsfig,graphics,subfigure,psfrag,amssymb}
\usepackage{lineno}
\usepackage{ulem}
\usepackage{dcolumn}
\usepackage{bm}
\usepackage{overpic}
\usepackage{xspace}
\usepackage{rotating}
\usepackage{color}
\usepackage{multirow}
\usepackage{colortbl}
\usepackage{epstopdf}
\definecolor{blackblue}{RGB}{17,118,208}
\definecolor{deepblue}{RGB}{0,0,51}
\usepackage[colorlinks,linkcolor=blackblue,anchorcolor=blue,citecolor=blackblue,urlcolor=blackblue]{hyperref}

\newcommand{\ee}{e^{+}e^{-}}

\newcommand{\BR}{{\cal B}}

\newcommand{\jpsi}{J/\psi}

\newcommand{\pim}{\pi^{-}}

\newcommand{\BESIII}{BES\uppercase\expandafter{\romannumeral3}\xspace}

\begin{document}
\title{%
    {Search for $J/\psi$ weak decays containing $D$ meson }}

    \author{M.~Ablikim$^{1}$, M.~N.~Achasov$^{5,b}$, P.~Adlarson$^{75}$, X.~C.~Ai$^{81}$, R.~Aliberti$^{36}$, A.~Amoroso$^{74A,74C}$, M.~R.~An$^{40}$, Q.~An$^{71,58}$, Y.~Bai$^{57}$, O.~Bakina$^{37}$
% ...
\\
\vspace{0.2cm}
(BESIII Collaboration)\\
\vspace{0.2cm} {\it
$^{1}$ Institute of High Energy Physics, Beijing 100049, People's Republic of China\\
% ...
$^{5}$ Budker Institute of Nuclear Physics SB RAS (BINP), Novosibirsk 630090, Russia\\
% ...
$^{36}$ Johannes Gutenberg University of Mainz, Johann-Joachim-Becher-Weg 45, D-55099 Mainz, Germany\\
$^{37}$ Joint Institute for Nuclear Research, 141980 Dubna, Moscow region, Russia\\
% ...
$^{40}$ Liaoning Normal University, Dalian 116029, People's Republic of China\\
% ...
$^{57}$ Southeast University, Nanjing 211100, People's Republic of China\\
$^{58}$ State Key Laboratory of Particle Detection and Electronics, Beijing 100049, Hefei 230026, People's Republic of China\\
% ...
$^{71}$ University of Science and Technology of China, Hefei 230026, People's Republic of China\\
% ...
$^{74}$ University of Turin and INFN, (A)University of Turin, I-10125, Turin, Italy; (B)University of Eastern Piedmont, I-15121, Alessandria, Italy; (C)INFN, I-10125, Turin, Italy\\
$^{75}$ Uppsala University, Box 516, SE-75120 Uppsala, Sweden\\
% ...
$^{81}$ Zhengzhou University, Zhengzhou 450001, People's Republic of China\\
\vspace{0.2cm}
% ...
$^{b}$ Also at the Novosibirsk State University, Novosibirsk, 630090, Russia\\
% ...
}
}


    \begin{abstract}
        Using a sample of about 10 billion $J/\psi$ events with the BESIII detector, we search for the weak decays of $J/\psi \to \bar{D}^0\pi^0 + c.c.$,  $J/\psi \to \bar{D}^0\eta + c.c.$, $J/\psi \to \bar{D}^0\rho^0 + c.c.$, $J/\psi \to D^-\pi^+ + c.c.$, and $J/\psi \to D^-\rho^+ + c.c.$. Since no significant signal is observed, we set the upper limits of the branching fractions of these decays to be $\mathcal{B}(J/\psi \to \bar{D}^0\pi^0 + c.c.) < 4.7 \times 10^{-7}$,  $\mathcal{B}(J/\psi \to \bar{D}^0\eta + c.c.) < 6.8 \times 10^{-7}$, $\mathcal{B}(J/\psi \to \bar{D}^0\rho^0 + c.c.) < 5.2 \times 10^{-7}$, $\mathcal{B}(J/\psi \to D^-\pi^+ + c.c.) < 7.0 \times 10^{-8}$, and $\mathcal{B}(J/\psi \to D^-\rho^+ + c.c.) <  6.0 \times 10^{-7}$ at the 90\% confidence level. 

    \end{abstract}

    \maketitle


    \section{Introduction}
    The $J/\psi$ meson is a bound state of charm quark and charm antiquark, with a mass of about 3.1 GeV/$c^2$~\cite{ParticleDataGroup:2022pth}. Lying below the threshold for the production of two open charm mesons, it cannot decay into two $D$ mesons. Its decays are dominated by strong and electromagnetic interactions, which have been extensively studied. Up to now, only a very limited number of rare weak decay channels have been studied experimentally~\cite{Ablikim:2006qt, Ablikim:2007ag, Ablikim:2014dsn, Ablikim:2014fpb, Ablikim:2017nid, BESIII:2021mnd}. Via the weak interaction, the $J/\psi$ can potentially decay into a single charm meson via such as $D$ accompanied by some non-charm mesons. Searching for the $J/\psi$ weak decays can provide an experimental test of the Standard Model (SM)~\cite{Chen:2021fcb}, which predicts the branching fractions of $J/\psi$ decays containing a $D$ meson up to an order of about $10^{-8}$~\cite{Verma:1990nk}. Furthermore, this search may offer a unique opportunity to probe new physics beyond the SM, including the top-color model~\cite{Hill:1994hp}, the minimal supersymmetric SM with or without R-parity violation~\cite{Aulakh:1982yn}, and the two-Higgs doublet model~\cite{Glashow:1976nt}, in which these branching fractions could be
    enhanced to be as large as $10^{-5}$~\cite{Datta:1998yq, Chen:2021fcb}.

\begin{figure}[htp]
        \centering
        \mbox{
            \begin{overpic}[width=4.3cm]{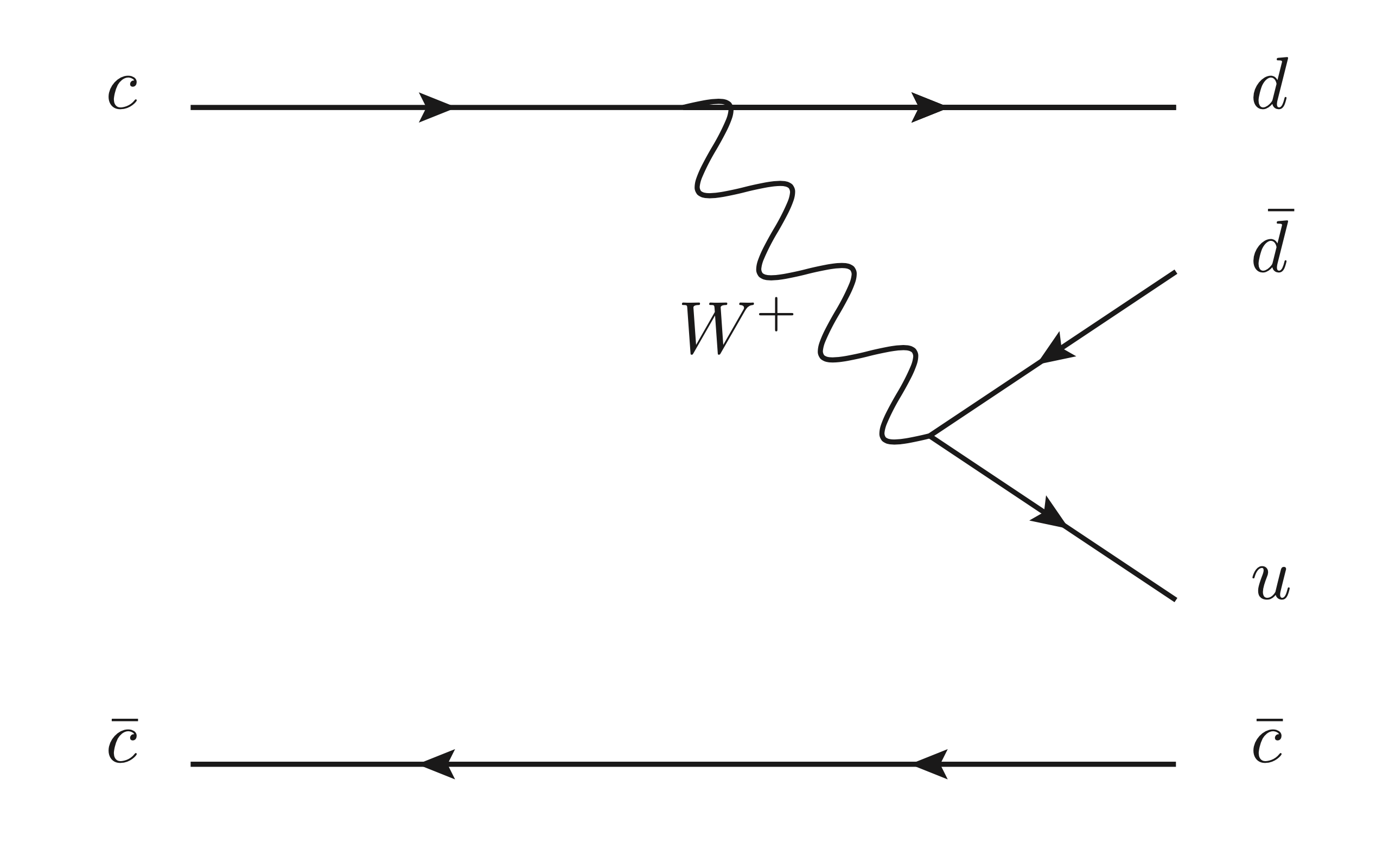}
                \put(40, -8){{\small (a)}}
            \end{overpic}
            \begin{overpic}[width=4.3cm]{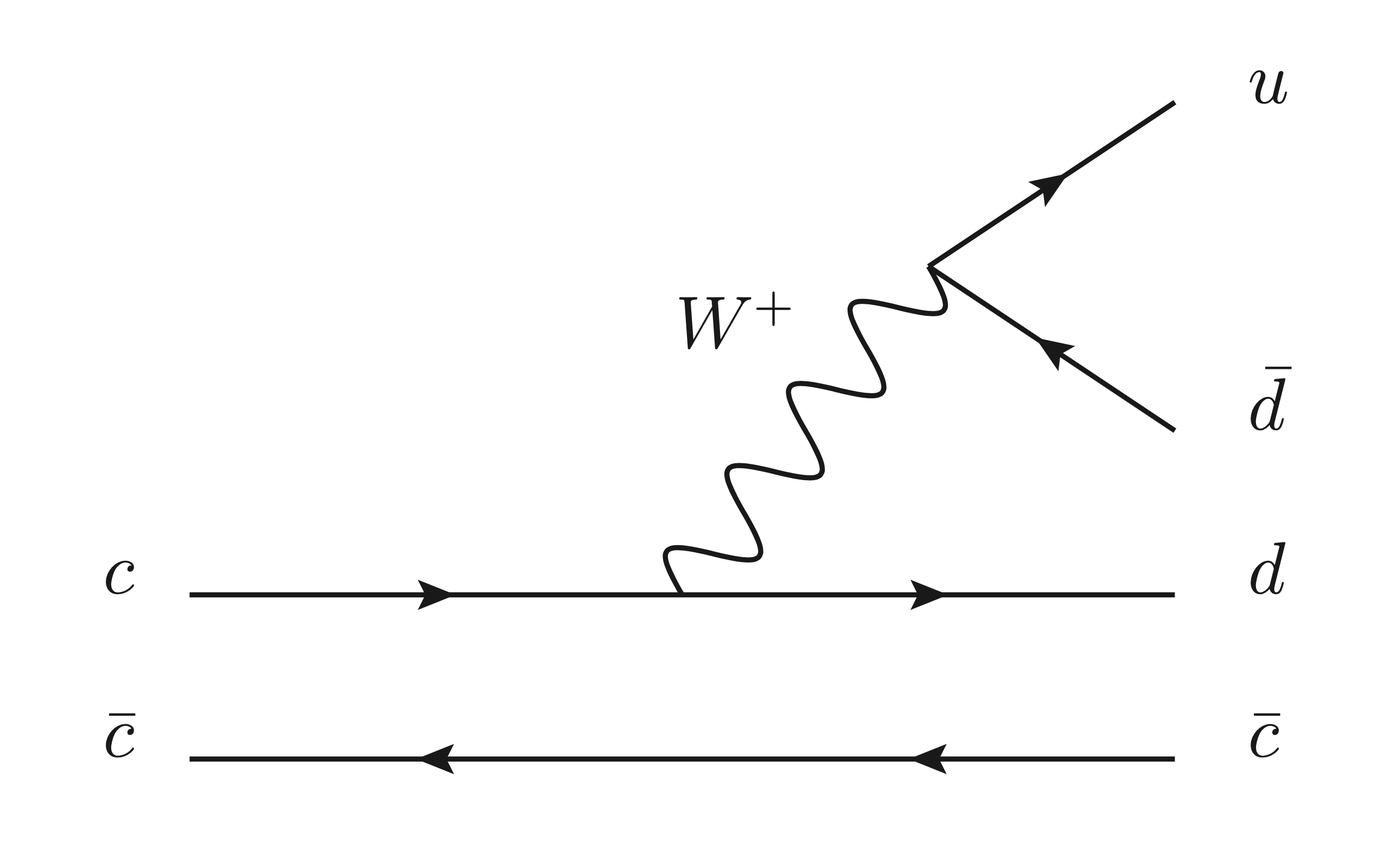}
                \put(40, -8){{\small (b)}}
            \end{overpic}
        }
        \caption{Leading-order Feynman diagrams of (a) $\jpsi \rightarrow \bar{D}^0\pi^{0}$, $\jpsi \rightarrow \bar{D}^0\eta$ and $\jpsi \rightarrow \bar{D}^0 \rho^0$; (b) $\jpsi \rightarrow D^-\pi^+$ and $\jpsi \rightarrow D^-\rho^+$.}\label{fig:fey}
    \end{figure}

    The weak hadronic decays of $\jpsi$ have been studied in theory, calculating the branching fractions for several decays of $J/\psi$ and $\psi(2S)$ into $(D_{(s)}+P)/(D_{(s)}+V)$, where $P$ and $V$ represent pseudoscalar mesons and vector mesons, respectively. From this, the ratio of the branching fractions was predicted at $\frac{\BR({\jpsi\to D_s^-\rho^+})}{\BR({\jpsi\to D_s^-\pim})} = 4.2$~\cite{Verma:1990nk, Sharma:1998gc}. Throughout this paper, charge conjugation is implied without specific indication. The $\jpsi\to D_{(s)}P$ decays, such as $\jpsi\to D^-\pi^{+}$, $\jpsi\to D_s^+\pi^-$ or $\jpsi\to D^0K^0$, were studied at BESII, and the upper limits on the branching fractions at the 90\% confidence level (C.L.) were set at the order of 10$^{-4}$, using a dataset of $5.8 \times 10^7$ $\jpsi$~\cite{Ablikim:2007ag}. For the $\jpsi \to D_{s} +V$ decay $\jpsi \to D_{s}^-\rho^{+}$, the upper limit on the branching fraction of this decay at the 90\% C.L. was determined to be of the order of 10$^{-5}$ with a data sample of 225.3 million $\jpsi$ events at BESIII~\cite{Ablikim:2014dsn}. However, for some $\jpsi \to DP$ and $J/\psi \to DV$ decays, such as $\jpsi\to \bar{D}^0\eta$, $\jpsi\to \bar{D}^0 \pi^0$, $\jpsi\to D^-\rho^+$, and $\jpsi \to \bar{D}^0 \rho^0$, which are mediated via $c\to d$ types, no experimental study has been reported so far. Figure~\ref{fig:fey} shows the Feynman diagrams for these decay modes in the SM. 

    Using a sample of  $(10087 \pm 44) \times 10^6 J/\psi$ events collected at the BESIII detector~\cite{BESIII:2021cxx}, we search for the weak decays $J/\psi \to \bar{D}^0\pi^0$,  $J/\psi \to \bar{D}^0\eta$, $J/\psi \to \bar{D}^0\rho^0$, $J/\psi \to D^-\pi^+$, and $J/\psi \to D^-\rho^+$. 

    \section{\texorpdfstring{Detectors and Data samples}{Detector and MC Simulation}}

    The BESIII detector~\cite{Ablikim:2009aa} records symmetric $e^+e^-$ collisions provided by the BEPCII storage ring~\cite{Yu:IPAC2016-TUYA01} in the center-of-mass energy range from 2.0 to 4.95~GeV with a peak luminosity of $1 \times 10^{33}\;\text{cm}^{-2}\text{s}^{-1}$ achieved at $\sqrt{s} = 3.77\;\text{GeV}$. BESIII has collected large data samples in this energy region~\cite{BESIII:2020nme}. The cylindrical core of the BESIII detector covers 93\% of the full solid angle and consists of a helium-based multilayer drift chamber~(MDC), a plastic scintillator time-of-flight system~(TOF), and a CsI(Tl) electromagnetic calorimeter~(EMC), which are all enclosed in a superconducting solenoidal magnet providing a 1.0~T (0.9~T in 2012) magnetic field. The solenoid is supported by an octagonal flux-return yoke with resistive plate counter muon identification modules interleaved with steel. 
    The charged-particle momentum resolution at $1~{\rm GeV}/c$ is $0.5\%$, and the ${\rm d}E/{\rm d}x$ resolution is $6\%$ for electrons from Bhabha scattering. The EMC measures photon energies with a resolution of $2.5\%$ ($5\%$) at $1$~GeV in the barrel (end cap) region. The time resolution in the TOF barrel region is 68~ps, while that in the end cap region was 110~ps. The end cap TOF system was upgraded in 2015 using multigap resistive plate chamber
    technology, providing a time resolution of 60~ps~\cite{etof}.

    Simulated data samples produced with a {\sc geant4}-based~\cite{geant4} Monte Carlo (MC) package, which includes the geometric description of the BESIII detector and the detector response~\cite{detvis}, are used to determine detection efficiencies and to estimate backgrounds. The simulation models the beam energy spread and initial state radiation (ISR) in the $e^+e^-$ annihilations with the generator {\sc kkmc}~\cite{ref:kkmc}. For the signal process, $\jpsi$ decays into D meson accompanied with a light hadron is generated using the QCDF decay model~\cite{Sun:2015nra}. The inclusive MC sample includes both the production of the $J/\psi$ resonance and the continuum processes incorporated in {\sc kkmc}~\cite{ref:kkmc}. All particle decays are modelled with {\sc evtgen}~\cite{Ping:2008zz, Lange:2001uf} using branching fractions either taken from the Particle Data Group~\cite{ParticleDataGroup:2022pth}, when available, or otherwise estimated with {\sc lundcharm}~\cite{Chen:2000tv}. Final state radiation~(FSR) from charged final state particles is incorporated using the {\sc photos} package~\cite{RichterWas:1992qb}.


    \section{Event selection and Data analysis}    
    To avoid high background from conventional $J/\psi$ hadronic decays, the $\bar{D}^0$ and $D^-$ mesons are tagged by the semileptonic decays $\bar{D}^0 \to K^+ e^- \bar{\nu}_e$ and $D^- \to K_{S}^{0} e^- \bar{\nu}_e$ with $K_{S}^{0} \to \pi^+ \pi^-$. Since the neutrino is undetectable at BESIII, the $\bar{D}^0$ and $D^-$ mesons cannot be directly reconstructed by the invariant mass of their decay products. However, for the two body $J/\psi$ decays investigated in this study, the $\bar{D}^0$ and $D^-$ mesons can be identified in the distributions of masses recoiling against the $\pi^0$, $\eta$, $\rho^0$, $\pi^+$, and $\rho^+$ with $\pi^0/\eta \to \gamma \gamma$, $\rho^0 \to \pi^+ \pi^-$, and $\rho^+ \to \pi^+ \pi^0$ decays, respectively. Specifically, for the signal decay modes $J/\psi \to \bar{D}^0\rho^0$ and $J/\psi \to D^-\rho^+$, to be conservative, we omit the non-$\rho$ contributions.

    Charged tracks detected in the MDC are required to be within a polar angle ($\theta$) range of $|\rm{cos\theta}|<0.93$, where $\theta$ is defined with respect to the $z$-axis, which is the symmetry axis of the MDC. For charged tracks, the distance of closest approach to the interaction point (IP) must be less than 10\,cm along the $z$-axis, $|V_{z}|$, and less than 1\,cm in the transverse plane, $|V_{xy}|$.

    Charged particle identification (PID) is performed by combining the TOF information and the ionization energy loss measured in the MDC. The information of the EMC is also included to identify electron candidates. Combined confidence levels for electron, pion and kaon hypotheses (CL$_e$, CL$_{\pi}$ and CL$_{K}$) are calculated individually. Charged tracks with CL$_{K(\pi)}>$ CL$_{\pi(K)}$ are identified as kaons (pions), and those with  CL$_e>$ CL{$_{\pi}$, CL$_e>$ CL$_{K}$ and CL$_e>0.001$ are identified as electrons.
    To further suppress the backgrounds from charged pions, the $E_{e}/p_{e} > 0.8$  requirement is imposed on electron candidates, where $E_{e}$ and $p_{e}$ are the deposited energy in the EMC and the momentum measured by the MDC, respectively.  

    Photon candidates are identified using showers in the EMC. The deposited energy of each shower must be more than 25~MeV in the barrel region ($|\cos \theta|< 0.80$) and more than 50~MeV in the end cap region ($0.86 <|\cos \theta|< 0.92$). To exclude showers that originate from charged tracks, the angle subtended by the EMC shower and the position of the closest charged track at the EMC must be greater than 10 degrees as measured from the IP. To suppress electronic noise and showers unrelated to the event, the difference between the EMC time and the event start time is required to be within [0, 700]\,ns.

    Each $K_{S}^0$ candidate is reconstructed from two oppositely charged tracks satisfying $|V_{z}|<$ 20~cm. The two charged tracks are assigned as $\pi^+\pi^-$ without imposing further PID criteria. They are constrained to
    originate from a common vertex and are required to have an invariant mass within $|M_{\pi^{+}\pi^{-}} - m_{K_{S}^{0}}|<$ 12~MeV$/c^{2}$, where $m_{K_{S}^{0}}$ is the $K^0_{S}$ nominal mass~\cite{ParticleDataGroup:2022pth}. The decay length of the $K^0_S$ candidate is required to be greater than twice the vertex resolution away from the IP. If there are multiple $K_{S}^{0}$ candidates in an event, the one with the smallest $\chi^2$ of the secondary vertex fit is retained.

    The $\pi^0/\eta$ candidates are reconstructed from candidate photon pairs. A kinematic fit, constraining the invariant mass of the photon pair to the world-average value of the $\pi^0/\eta$ mass~\cite{ParticleDataGroup:2022pth} is performed. The combination with the minimum $\chi^2$ from the kinematic fit and satisfying $\chi^2 < 20$ and $0.115<M(\gamma\gamma)<0.150$ GeV/$c^2$ ($0.50<M(\gamma\gamma)<0.57$ GeV/$c^2$) for $\pi^0$ ($\eta$) is kept for further analysis. The $\rho^0$ and $\rho^+$ candidates are selected in the regions $0.62 < M_{\pi^+\pi^-}/M_{\pi^+\pi^0} < 0.95$ GeV/c$^2$. 

    The numbers of charged track candidates are two, two, four, four, and four, while at least two, two, zero, zero, and two photons are required for $J/\psi \to \bar{D}^0\pi^{0}$, $J/\psi \to \bar{D}^0\eta$, $J/\psi \to \bar{D}^0\rho^0$, $J/\psi \to D^-\pi^+$, and $J/\psi \to D^-\rho^+$, respectively.

    \begin{figure*}[htp]
        \centering
        \mbox{
            \subfigure{
                \label{UmissA}
                \begin{overpic}[width=5.5cm]{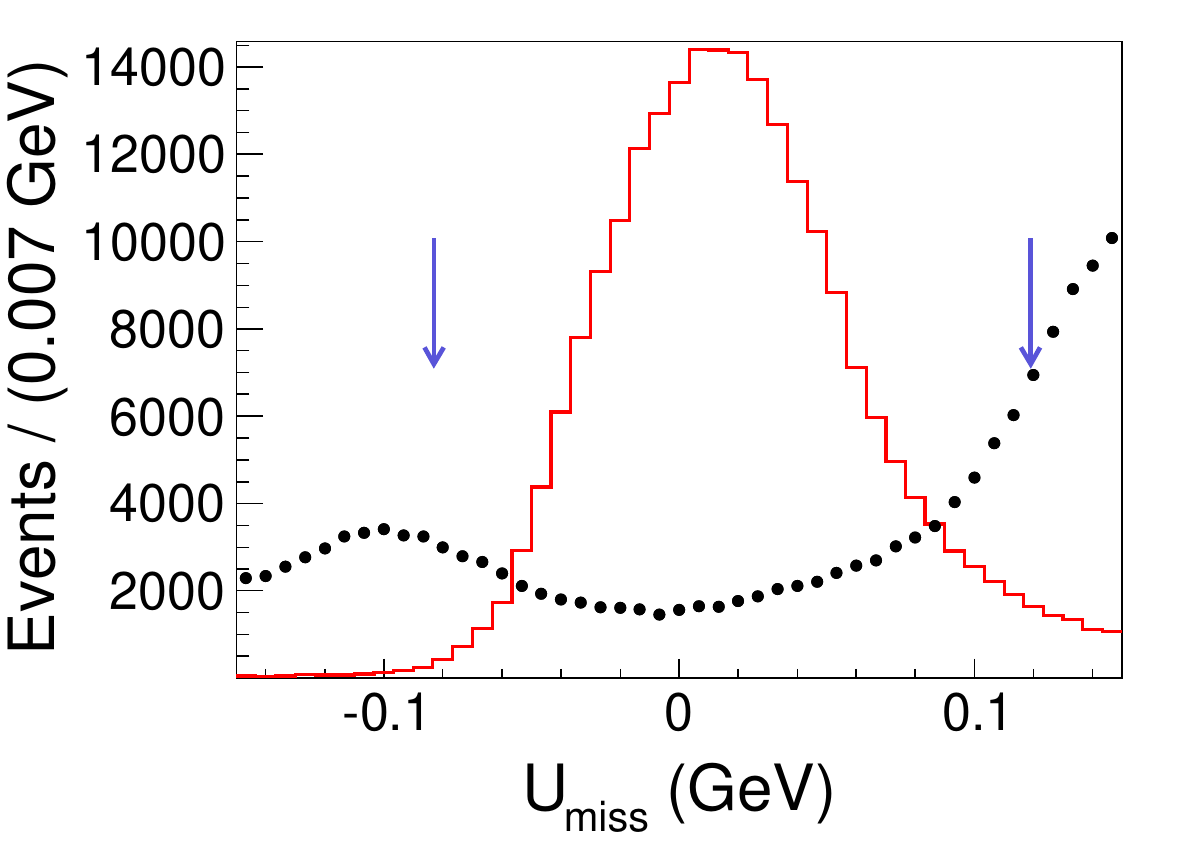}
                    \put(30,60){{\small (a)}}
                \end{overpic}
            }
            \subfigure{
                \label{UmissB}
                \begin{overpic}[width=5.5cm]{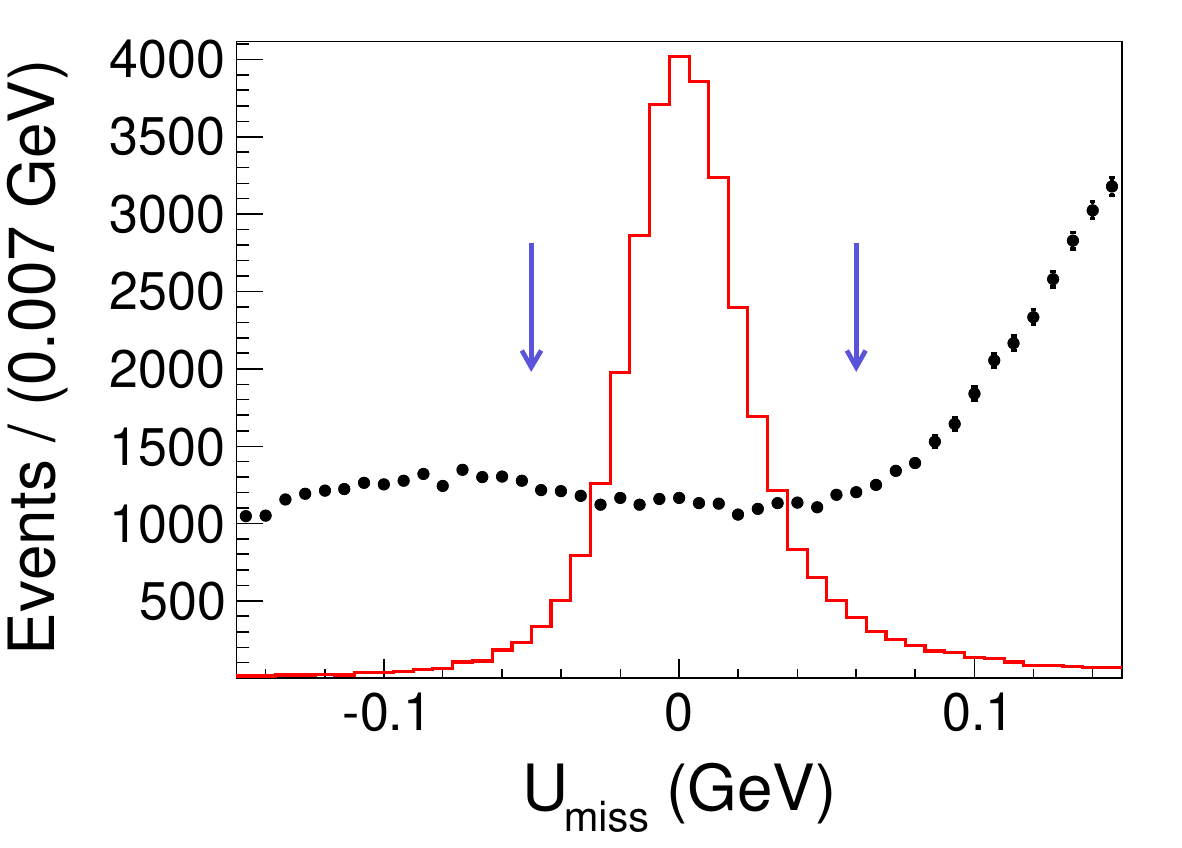}
                    \put(30,60){{\small(b)}}
                \end{overpic}
            }
        }
        \mbox{
            \subfigure{
                \label{UmissC}
                \begin{overpic}[width=5.5cm]{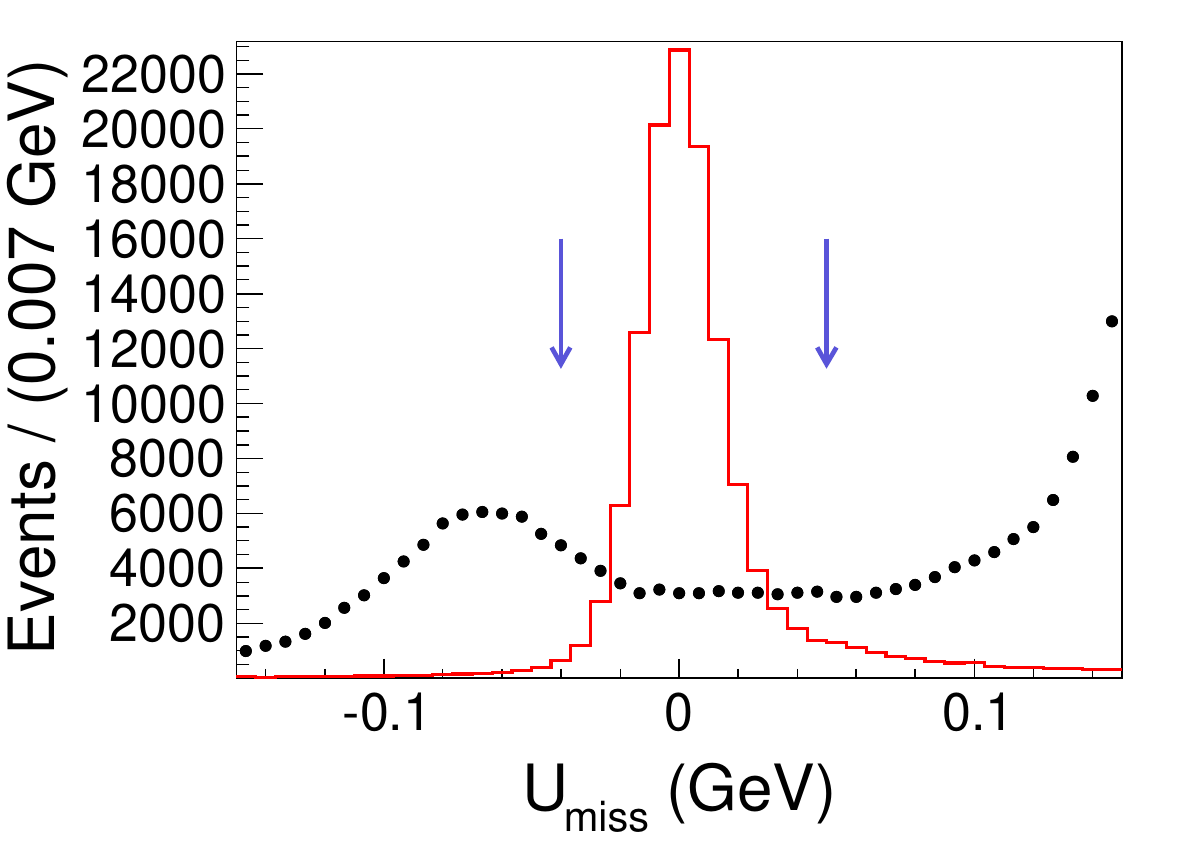}
                    \put(30,60){{\small (c)}}
                \end{overpic}
            }
            \subfigure{
                \label{UmissD}
                \begin{overpic}[width=5.5cm]{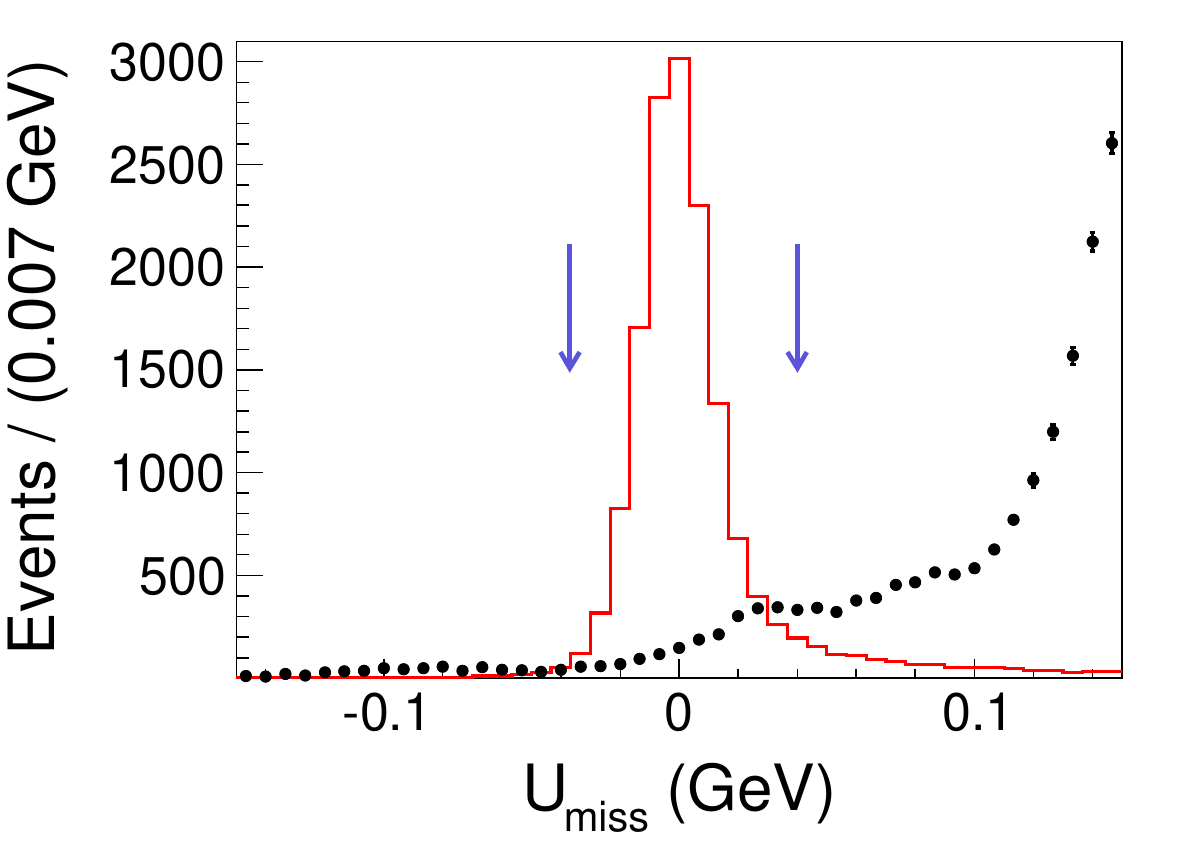}
                    \put(30,60){{\small (d)}}
                \end{overpic}
            }
        }
        \subfigure{
            \label{UmissE}
            \begin{overpic}[width=5.5cm]{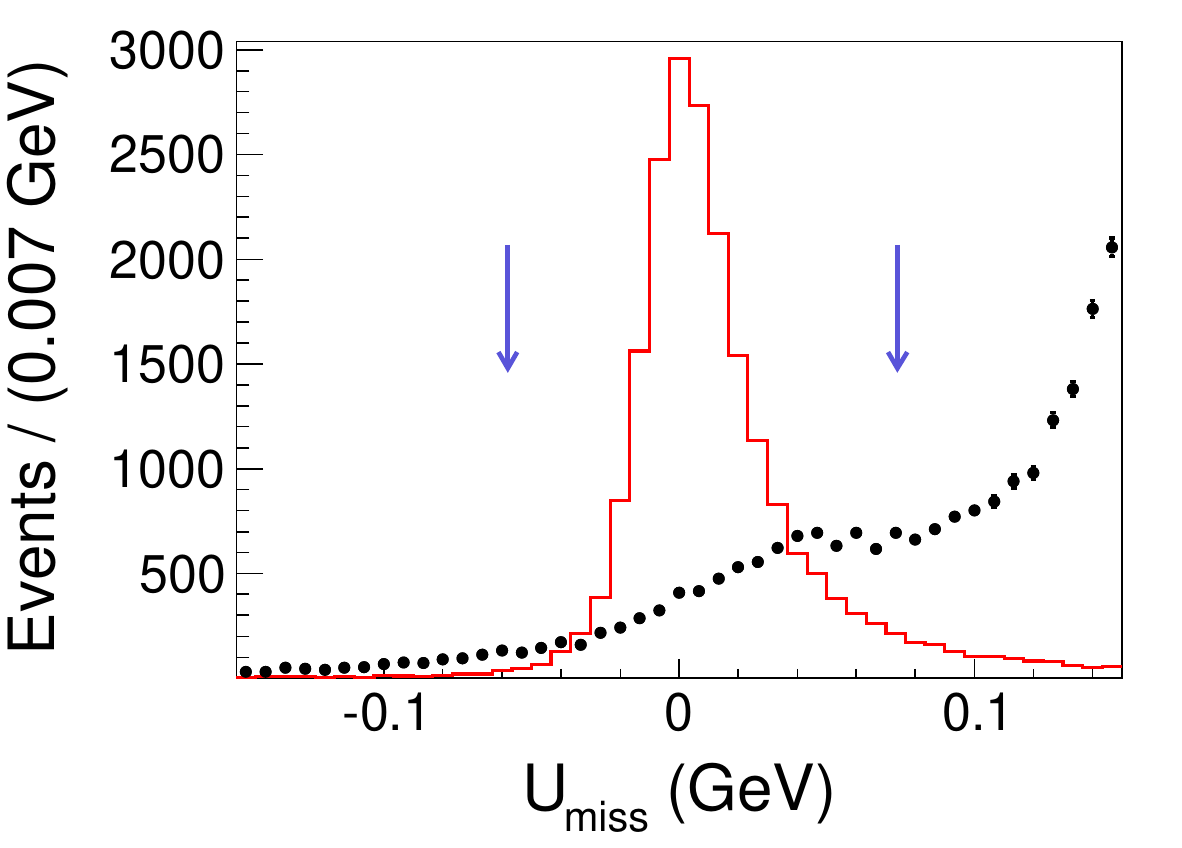}
                \put(30,60){{\small (e)}}
            \end{overpic}
        }       
        \caption{Distribution of $U_{\rm miss}$ from (a) $\jpsi\to\bar{D}^0\pi^0$, (b) $J/\psi \to \bar{D}^0\eta$, (c) $J/\psi \to \bar{D}^0\rho^0$, (d) $J/\psi \to D^-\pi^+$, and (e) $J/\psi \to D^-\rho^+$. The black dots with error bars represent data and the red thick lines show the signal MC sample. The region between the two blue arrows marks the signal region of $U_{\rm miss}$.}
        \label{fig:Umiss}
    \end{figure*}

    For the channel $J/\psi \to \bar{D}^0 M$, where $M$ represents the mesons $\pi^{0}$, $\eta$, or $\rho^0$, the undetected neutrino carries a missing-energy $E_{\rm miss} = E_{J/\psi} - E_{M} - E_{K^{+}} - E_{e^{-}}$ and a missing-momentum $\vec{p}_{\rm miss} = \vec{p}_{J/\psi} - \vec{p}_{M} - \vec{p}_{K^{+}} - \vec{p}_{e^{-}}$ according to energy-momentum conservation. Correspondingly, the missing-energy and missing-momentum for the decay mode $J/\psi \to {D}^- N$, where $N$ marks the mesons $\pi^+$ and $\rho^+$, are $E_{J/\psi} - E_{N} - E_{K_{S}^{0}} - E_{e^{-}}$ and $\vec{p}_{J/\psi} - \vec{p}_{N} - \vec{p}_{K_{S}^{0}} - \vec{p}_{e^{-}}$. Here, the energies and momenta of $M$, $N$, $K^{+}$, $K_{S}^{0}$, and $e^{-}$ are taken in the rest frame of the initial $e^+e^-$ collision. The kinematic quantity $U_{\rm miss}=E_{\rm miss}-c|\vec{p}_{\rm miss}|$ is used to identify the missing neutrino and the criterion of $U_{\rm miss}$ is applied to suppress the backgrounds with multi-$\pi^0/ \gamma$ and the misidentification of electron/pion and kaon/pion in the final states. The requirements of $U_{\rm miss}$ for the decay modes $J/\psi \to \bar{D}^0\pi^{0}$, $J/\psi \to \bar{D}^0\eta$ , $J/\psi \to \bar{D}^0\rho^0$, $J/\psi \to D^-\pi^+$, and $J/\psi \to D^-\rho^+$  are within the regions $(-0.083, 0.119)$, $(- 0.050,0.060)$, $(-0.040, 0.050)$, $(-0.037,0.040)$, and $(-0.058,0.074)$~GeV, respectively. Figure~\ref{fig:Umiss} shows the distributions of  $U_{\rm miss}$ of the accepted candidates for the five decay modes. From the inclusive MC sample, no obvious peaking background in the signal regions is observed. We select those events for which the recoiling mass against the $\pi^0$, $\eta$, $\rho^0$, $\pi^+$, and $\rho^+$ falls within the mass window (1.80, 1.95) GeV/$c^2$ for all decay modes. Using signal MC events,  the detection efficiencies for $J/\psi \to \bar{D}^0\pi^{0}$, $J/\psi \to \bar{D}^0\eta$ , $J/\psi \to \bar{D}^0\rho^0$, $J/\psi \to D^-\pi^+$, and $J/\psi \to D^-\rho^+$ are determined to be 41.3\%, 34.2\%, 32.2\%, 35.5\%, and 14.2\%, respectively.


    \section{Upper Limits}

    Figure~\ref{fig:RmassFit} shows the recoiling mass spectra of the accepted candidates for $\jpsi\to\bar{D}^0\pi^0$, $J/\psi \to \bar{D}^0\eta$, $J/\psi \to \bar{D}^0\rho^0$, $J/\psi \to D^-\pi^+$, and $J/\psi \to D^-\rho^+$. No significant signal is observed in any of the decay modes. As shown in Fig.~\ref{fig:RmassFit}, an unbinned extended maximum likelihood fit is performed to extract the signal yields. In the fits, the signal is modeled by the signal MC shape of the recoiling mass spectrum and the background is modeled by a first-order polynomial function. Table~\ref{tab:results} shows the fit results. The branching fraction of signal decay is calculated as
    \begin{equation}
        \mathcal{B}(\jpsi \to DM(N)) = \frac{N_{\rm sig}}{N_{\jpsi} \times \epsilon \times \mathcal{B}_{\rm sub}}, 
    \end{equation}
    where $N_{\rm sig}$ is the number of signal events, $N_{\jpsi}$ is the total number of $\jpsi$ events~\cite{BESIII:2021cxx}, $\epsilon$ is the signal detection efficiency, and $\mathcal{B}_{\rm sub}$ is the product of the branching fractions of all possible intermediate decays.

    \begin{figure*}[htp]
        \centering
        \mbox{
            \begin{overpic}[width=5.5cm]{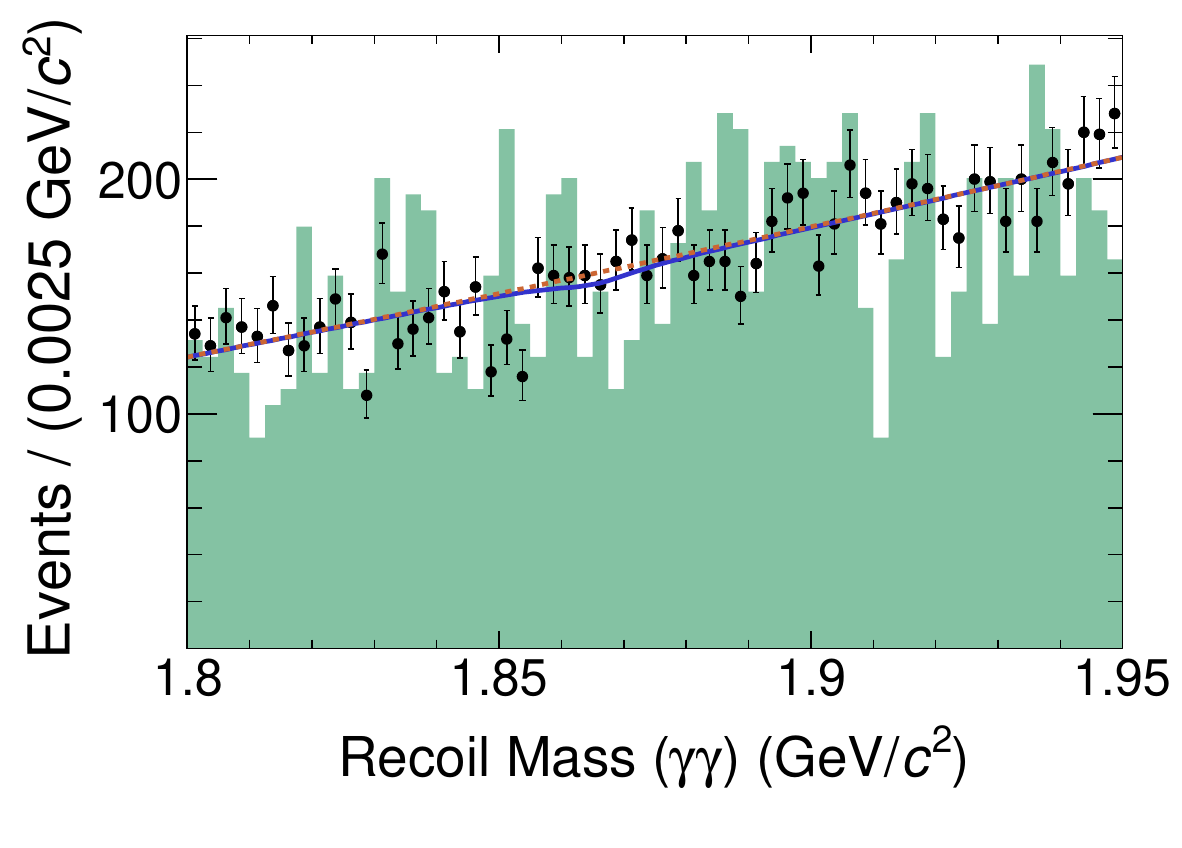}
                \put(20,60){{\small (a)}}
            \end{overpic}

            \begin{overpic}[width=5.5cm]{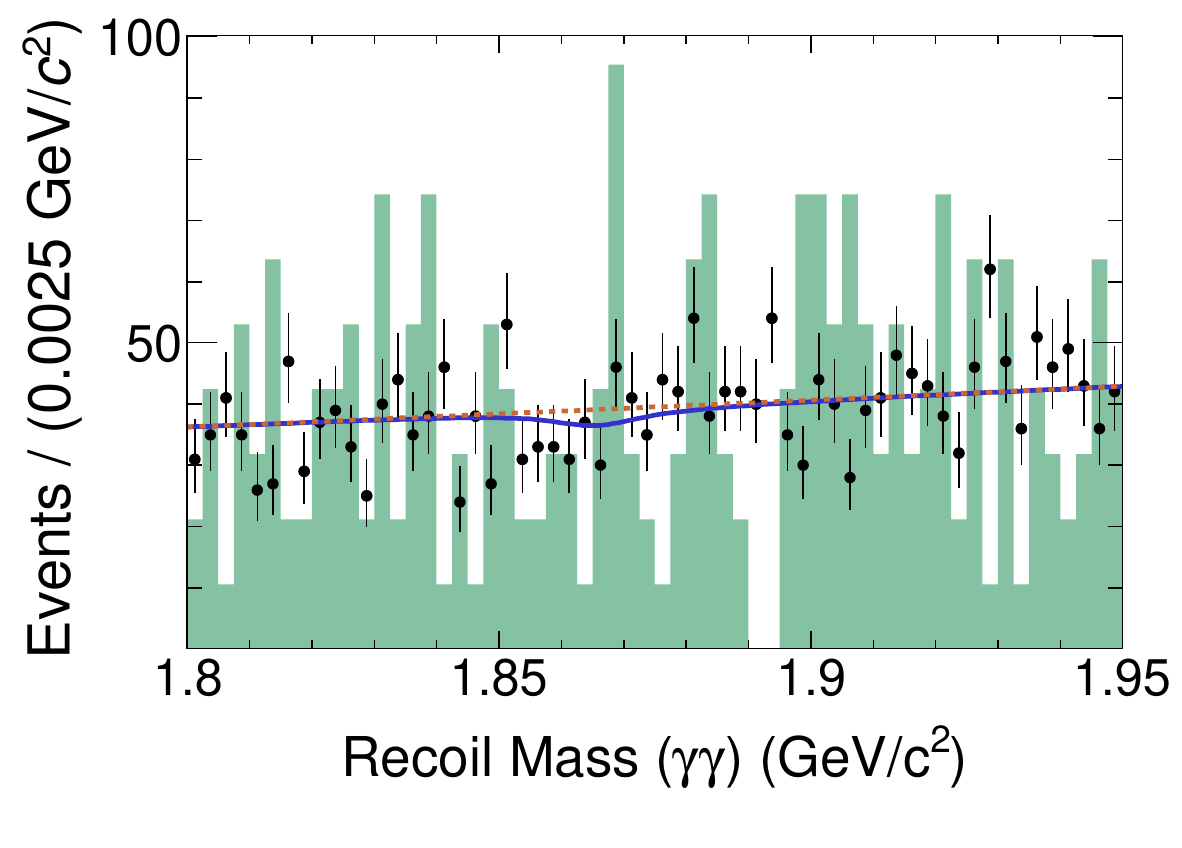}
                \put(20,60){{\small (b)}}
            \end{overpic}
        }

        \mbox{  
        \begin{overpic}[width=5.5cm]{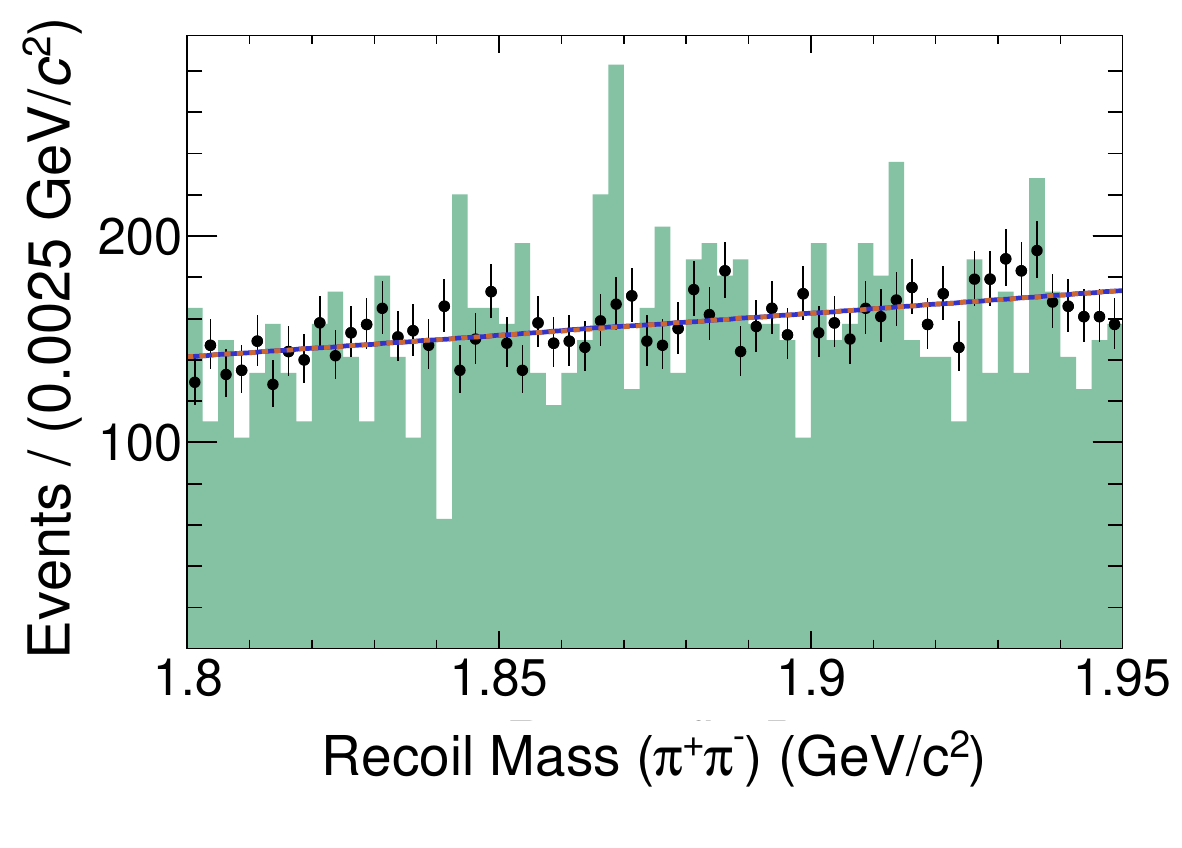}
            \put(20,60){{\small (c)}}
        \end{overpic}

        \begin{overpic}[width=5.5cm]{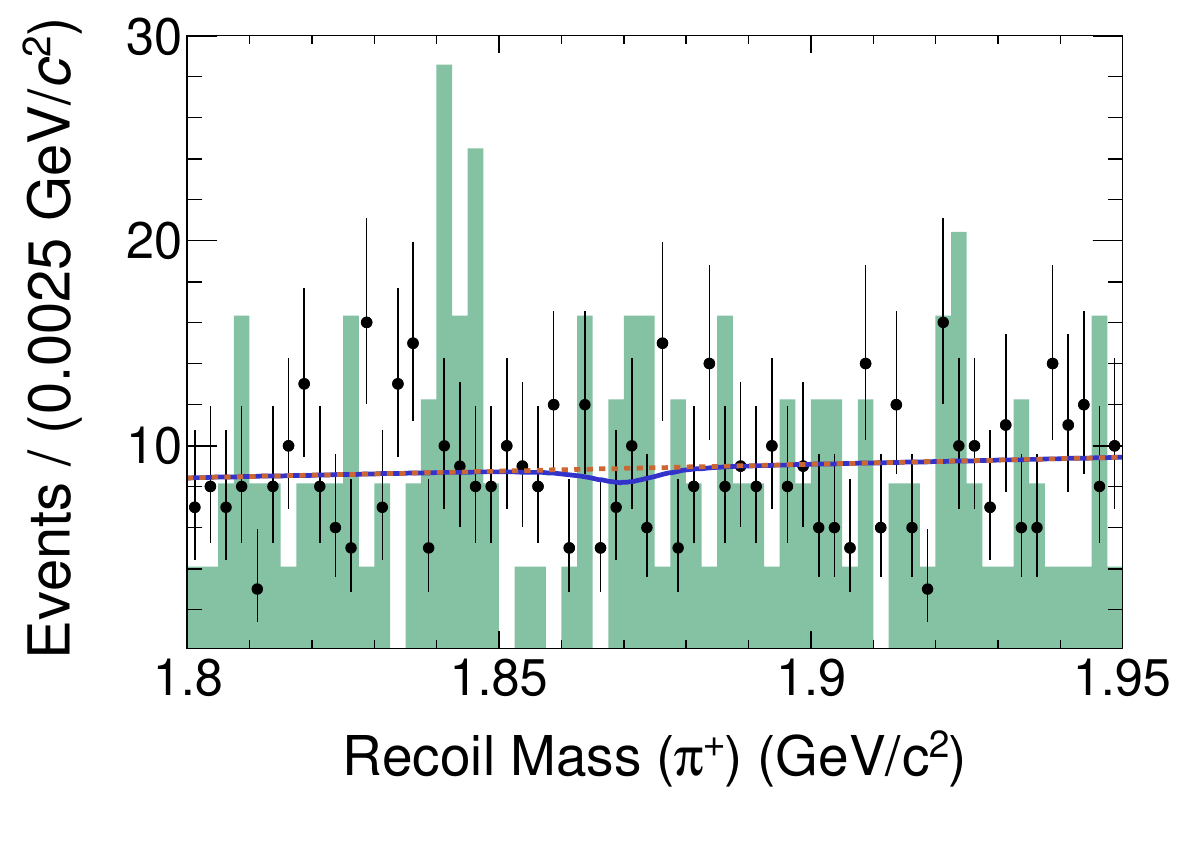}
            \put(20,60){{\small (d)}}
        \end{overpic}
        }

        \begin{overpic}[width=5.5cm]{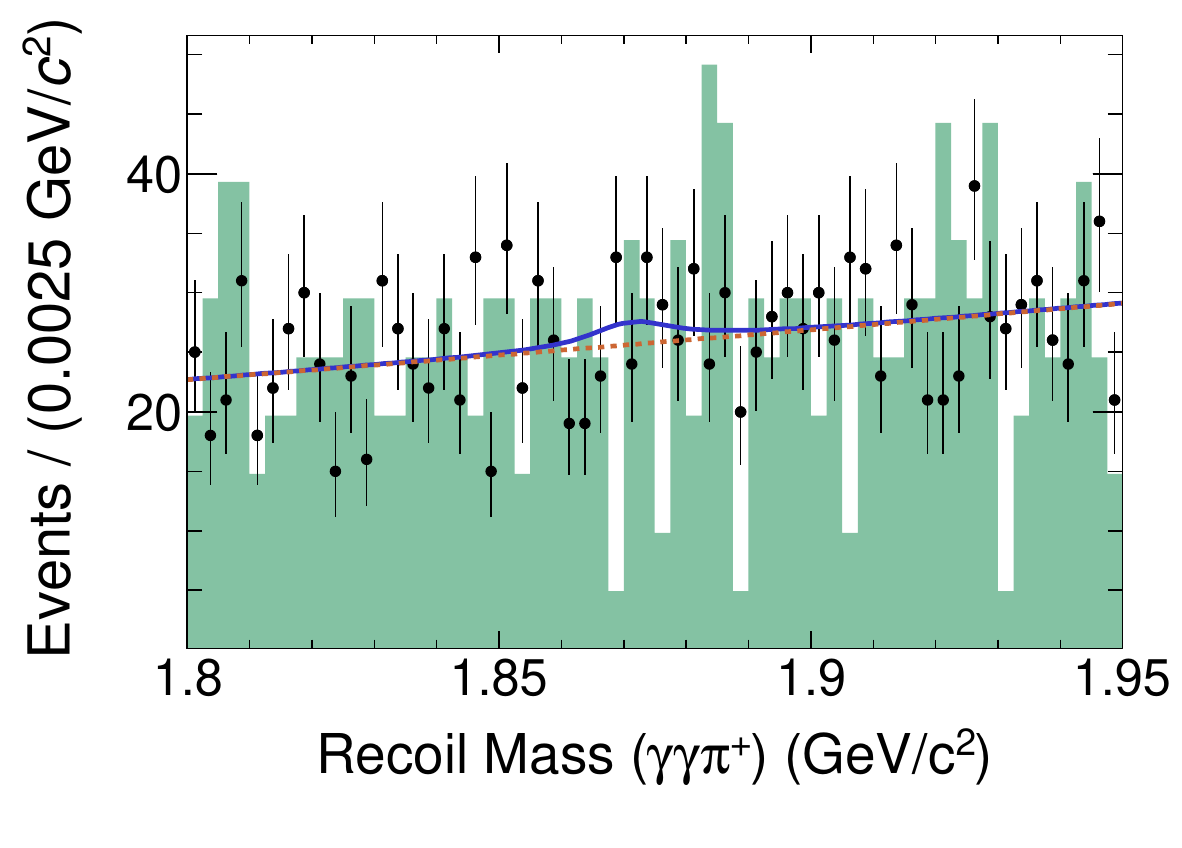}
            \put(20,60){{\small (e)}}
        \end{overpic}

        \caption{Fits of the accepted candidates to the recoiling mass spectra for (a) $\jpsi\to\bar{D}^0\pi^0$, (b) $J/\psi \to \bar{D}^0\eta$, (c) $J/\psi \to \bar{D}^0\rho^0$, (d) $J/\psi \to D^-\pi^+$, and (e) $J/\psi \to D^-\rho^+$. The dots with error bars are data and the orange dotted lines are polynomial functions describing the background. The blue solid curves are the total fits. The inclusive MC samples are shown by the green filled histograms.}
        \label{fig:RmassFit}
    \end{figure*}

    To set the upper limit on the branching fraction via a Bayesian approach~\cite{bernardo:2000}, we perform a likelihood scan with a series of fits, where the numbers of signal events $N_{\rm sig}$ are fixed to a series of values in the scan region, which are shown in Table~\ref{tab:results}. Since the branching fraction is only meaningful in the physical region ($\mathcal{B}\geq0$), the upper limit on the branching fraction is calculated in this region by taking into account the systematic uncertainties, which include additive and multiplicative items as described in Sec.~V. The additive uncertainties are irrelevant to the efficiencies but are associated with the fit procedure, so they are considered separately. We repeat the maximum-likelihood fits by varying the background shape and take the most conservative upper limit among different choices of background shapes. Then we follow the method discussed in Ref.~\cite{uplimit} that incorporates multiplicative systematic uncertainties into the upper limits. The distribution of the maximum likelihood scan $L(n)$, as a function of the yield $n$ is smeared with the multiplicative uncertainty $\sigma_{\epsilon}$, which is the quadratic sum of the various multiplicative systematic uncertainties, namely    \begin{equation}
        L(n)\propto \int^{1}_{0}L\left(n\frac{\epsilon}{\epsilon_{0}}\right) {\rm exp}\left[\frac{-(\epsilon/\epsilon_{0} - 1)^{2}}{2\sigma_{\epsilon}^2}\right] {\rm d}\epsilon,
    \end{equation}
    where $\epsilon_{0}$ is the nominal efficiency based on the signal MC sample. The normalized likelihood versus $N_{\rm sig}$ is shown in Fig.~\ref{fig:likelihood}, and the upper limits on the branching fractions at the $90\%$ C.L. are obtained by integrating from zero to $90\%$ of the likelihood curve in the physical region. The results are summarized in Table~\ref{tab:results}.

    \begin{table*}
        \centering
        \caption{The signal yields $N_{\rm sig}$ obtained from fits and the upper limits on the signal yields $N^{\rm UL}_{\rm sig}$ and branching fractions ${\mathcal B}$ at the 90\% C.L., where the uncertainties of $N_{\rm sig}$ are statistical only, and the fifth column represents the previous results.
        \label{tab:results} }
        \begin{tabular}{lcllcc}\hline\hline
            Mode                                  &\qquad                                        &$N_{\rm sig}$         &$N^{\rm UL}_{\rm sig}$    &${\mathcal B}$ (90\% C.L.)     &${\mathcal B}$ (90\% C.L.)   \\ \hline
            $\jpsi\to \bar{D}^0\pi^0$   &  &$-49.5\pm69.3$          &$< 68.8$            &$< 4.7 \times 10^{-7}$   &$\cdots$ \\
            $\jpsi\to \bar{D}^0\eta$     &   &$-28.9\pm34.5$          &$< 32.9$             &$< 6.8 \times 10^{-7}$  &$\cdots$ \\
            $\jpsi\to \bar{D}^0\rho^0$  &  &$2.0\pm37.1$          &$< 59.9$            &$< 5.2 \times 10^{-7}$ &$\cdots$\\
            $\jpsi\to D^-\pi^+$              &  &$-4.3\pm10.3$         &$< 14.4$           &$< 7.0 \times 10^{-8}$  & $< 7.5 \times 10^{-5}$~\cite{Ablikim:2007ag}\\
            $\jpsi\to D^-\rho^+$            &   &$18.6\pm26.2$       &$< 51.4$            &$< 6.0 \times 10^{-7}$    &$\cdots$ \\ \hline\hline
        \end{tabular}
    \end{table*}

    \begin{figure*}[htp]
        \centering
        \mbox{
            \subfigure{
                \label{UppLimA}
                \begin{overpic}[width=5.5cm]{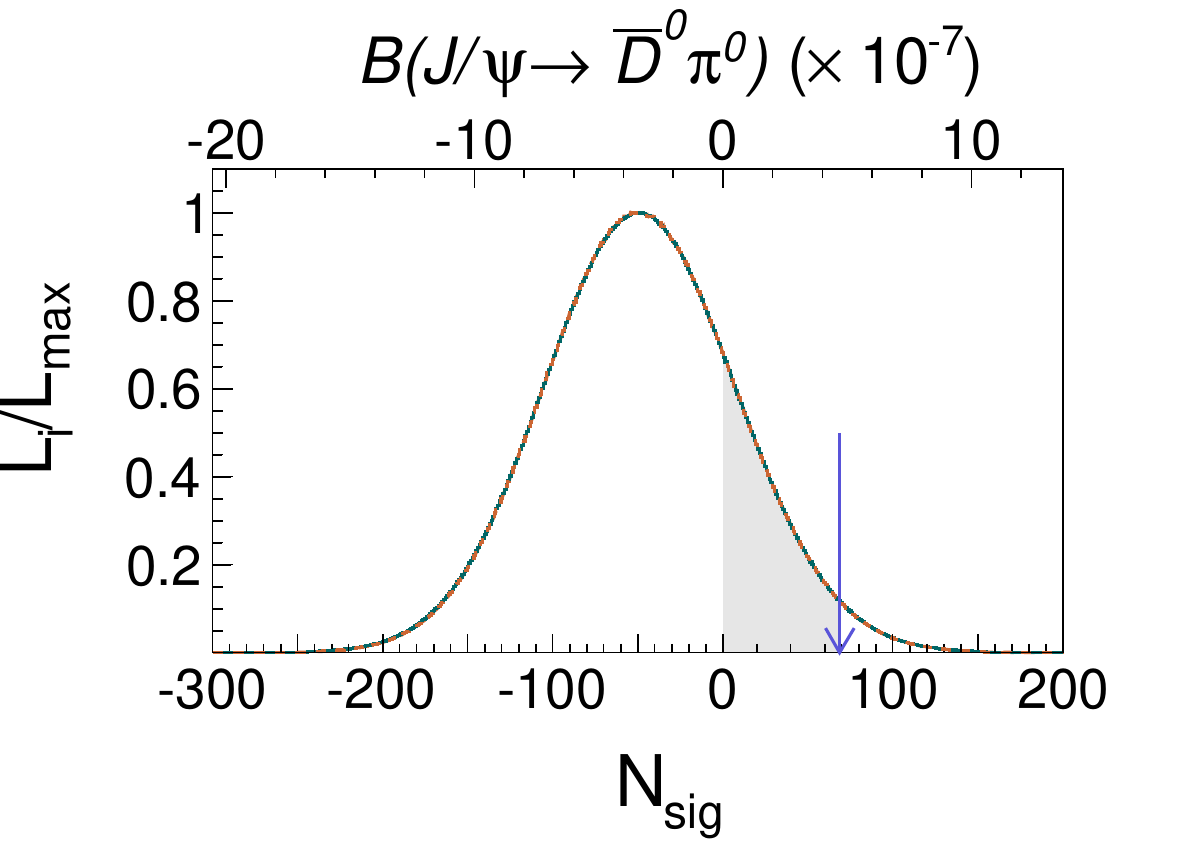}
                    \put(25,50){{\small (a)}}
                \end{overpic}
            }
            \subfigure{
                \label{UppLimB}
                \begin{overpic}[width=5.5cm]{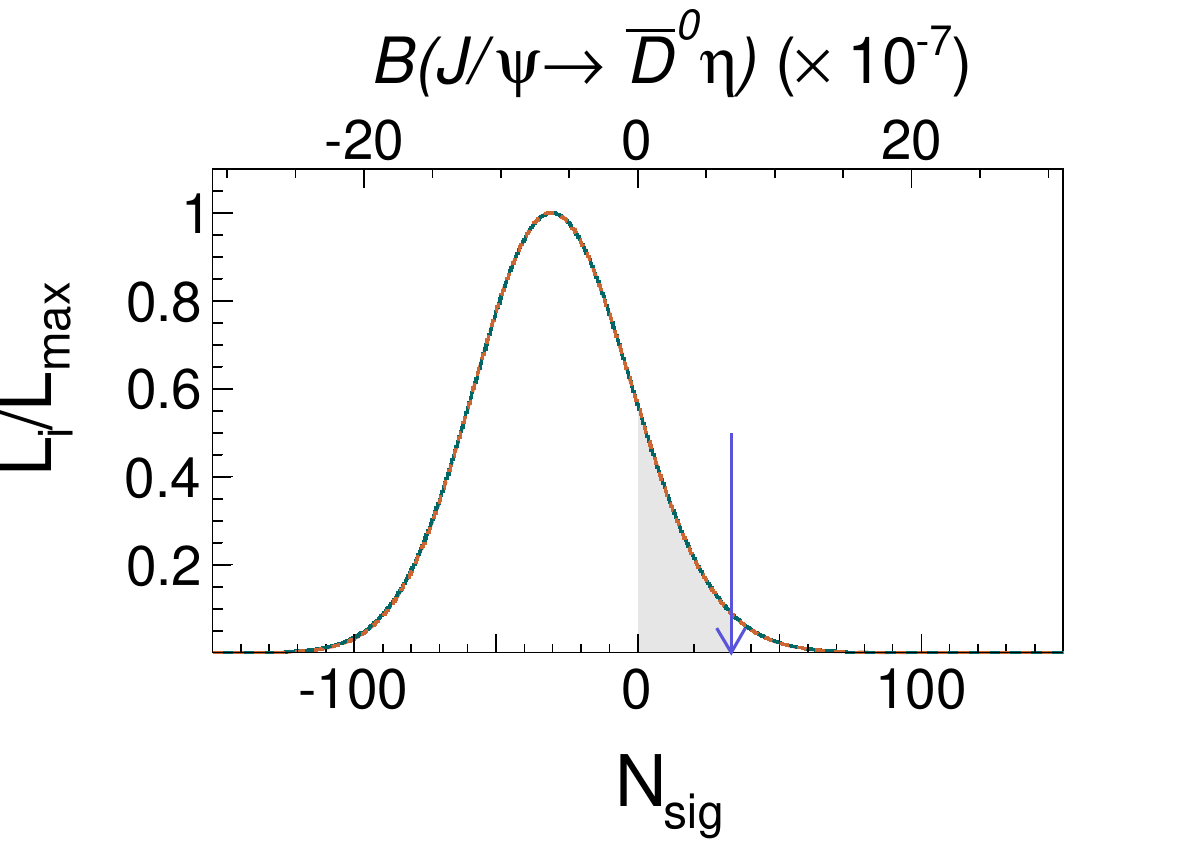}
                    \put(25,50){{\small (b)}}
                \end{overpic}
            }
        }

        \mbox{
            \subfigure{
                \label{UppLimC}
                \begin{overpic}[width=5.5cm]{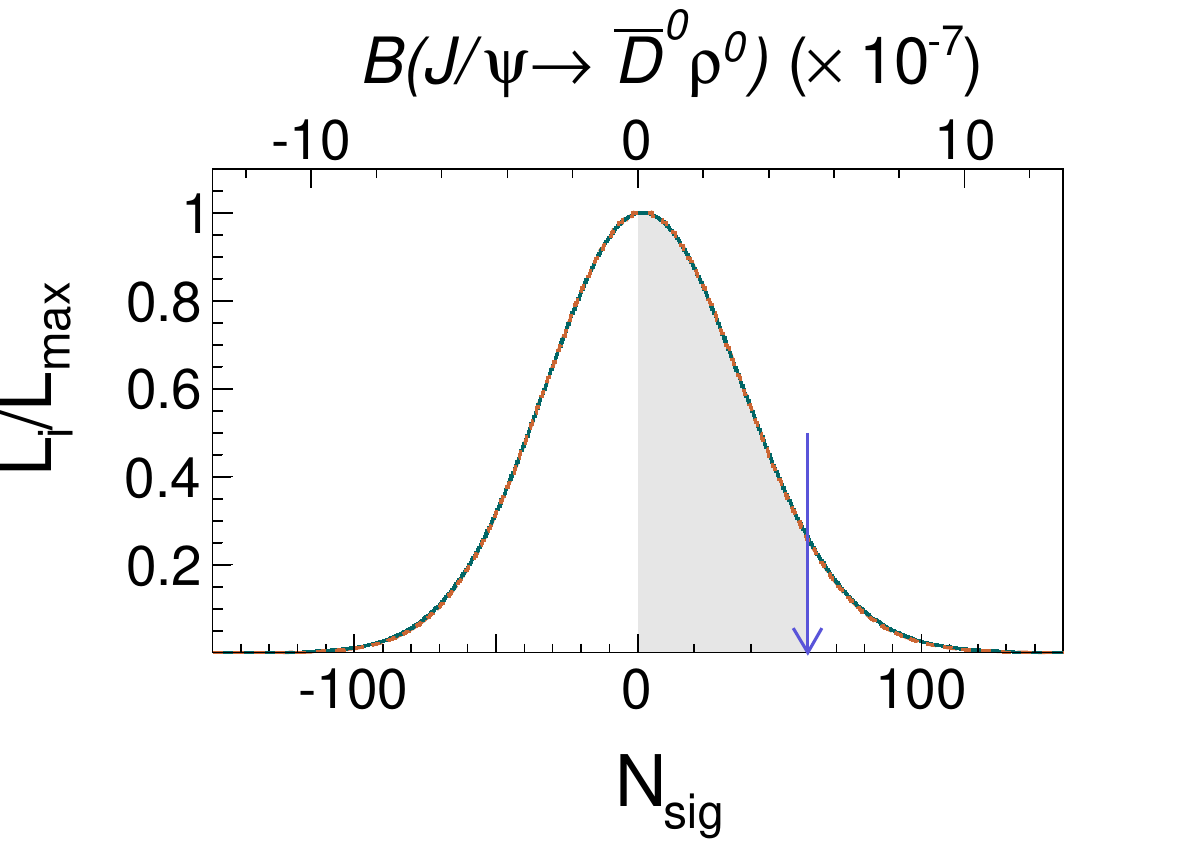}
                    \put(25,50){{\small (c)}}
                \end{overpic}
            }
            \subfigure{
                \label{UppLimD}
                \begin{overpic}[width=5.5cm]{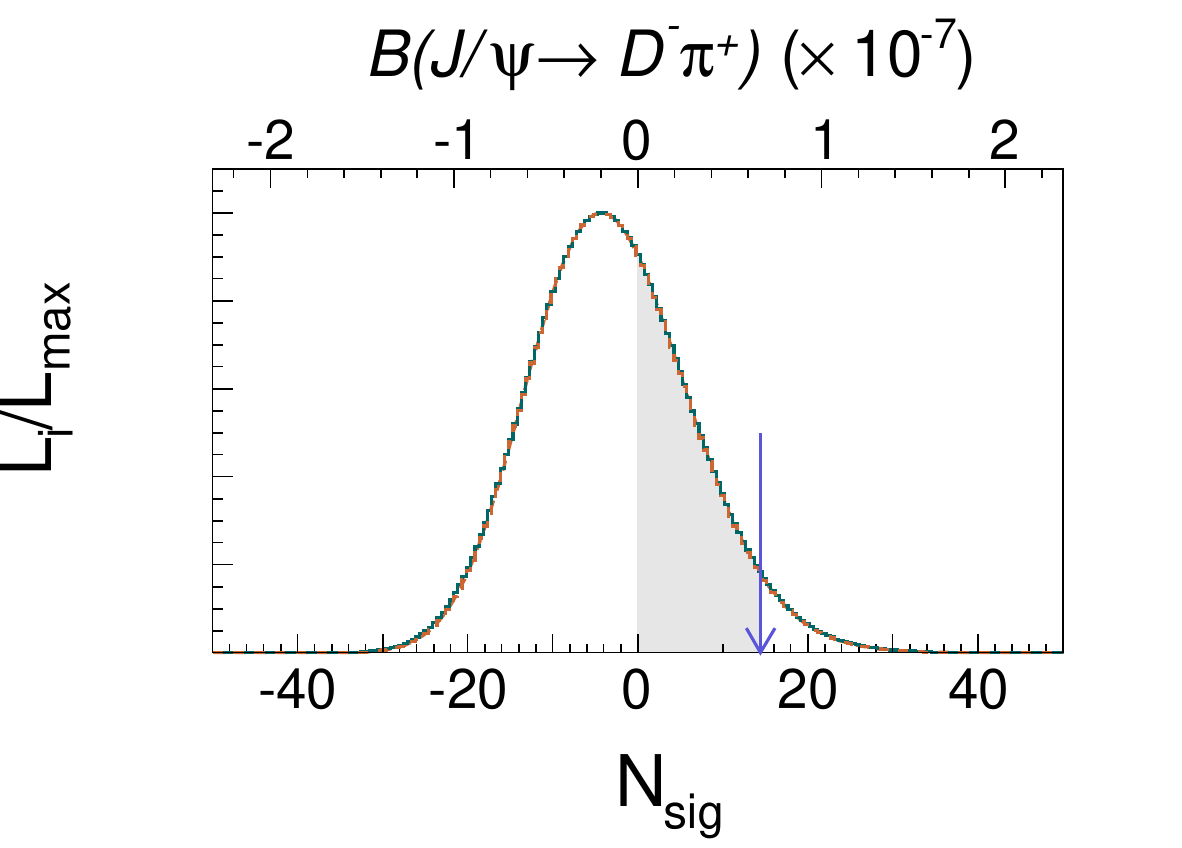}
                    \put(25,50){{\small (d)}}
                \end{overpic}
            }
        }
        \subfigure{
            \label{UppLimE}
            \begin{overpic}[width=5.5cm]{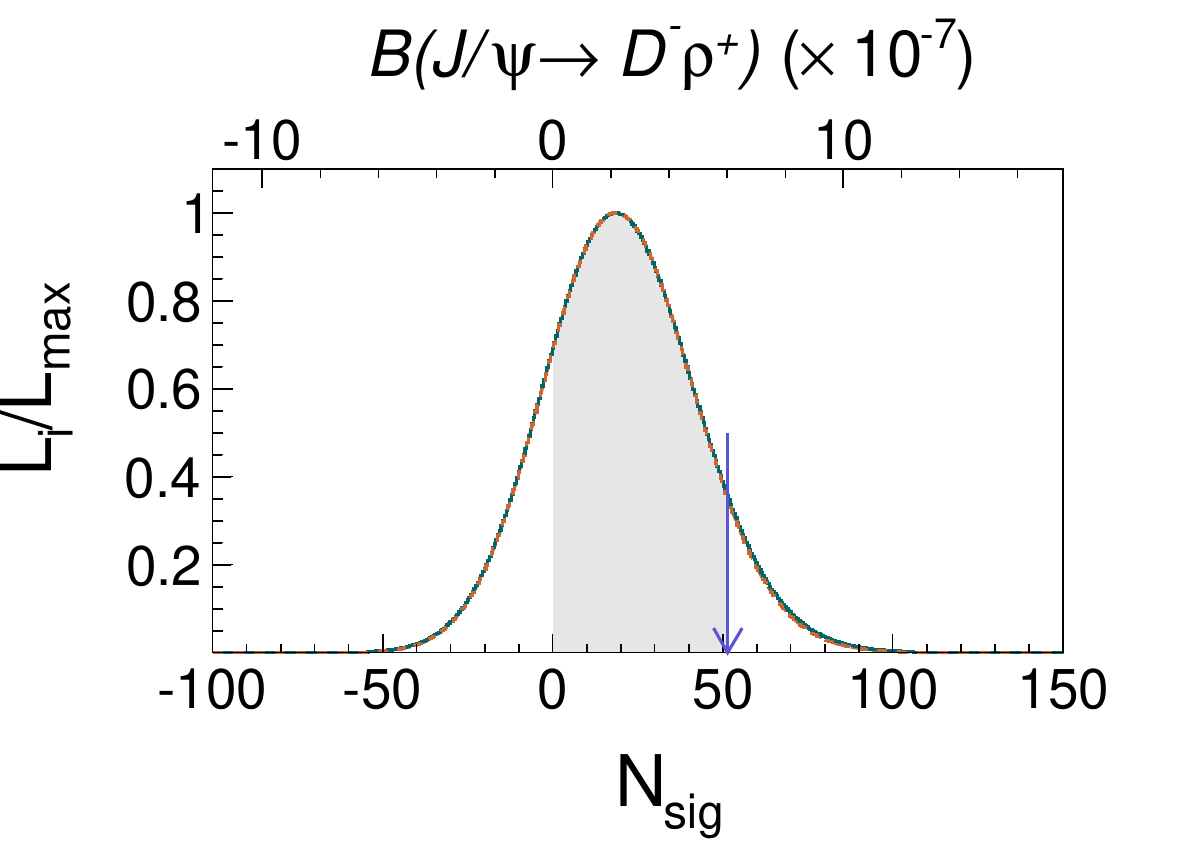}
                \put(25,50){{\small (e)}}
            \end{overpic}
        }

        \caption{Normalized likelihood distributions for the fitted yields of signal events and corresponding branching fractions of (a) $\jpsi\to\bar{D}^0\pi^0$, (b) $\jpsi\to\bar{D}^0\eta$, (c) $\jpsi\to\bar{D}^0\rho^0$, (d) $\jpsi\to D^-\pi^+$, and (e) $\jpsi\to D^-\rho^+$, with (green solid curves) and without (orange dashed lines) smearing the systematic uncertainties. The blue arrows mark the upper limits at the 90\% C.L.}
        \label{fig:likelihood}
    \end{figure*}


    \section{Systematic uncertainties}

    \begin{table*}
        \centering
        \caption{Multiplicative systematic uncertainties in the measured branching fractions for  $\jpsi \to \bar{D}^0 \pi^0$, $\bar{D}^0 \eta$, $\bar{D}^0 \rho^0$, $D^- \pi^+$, and $D^- \rho^+$.\label{sys:tot}}
        \scalebox{1.}{
            \begin{tabular}{cccccccc}\hline\hline
                \multicolumn{6}{c}{Multiplicative (in \%)}  \\  \hline
                Source  & $\jpsi \to \bar{D}^0 \pi^0$   & $\jpsi \to \bar{D}^0 \eta$         & $\jpsi \to \bar{D}^0 \rho^0$  &$\jpsi \to D^- \pi^+$  & $\jpsi \to D^- \rho^+$         \\
                Tracking   			  &2.0 	                       & 2.0                                    &4.0   	                         & 4.0              & 4.0			                 \\
                PID           		          &2.0		                   & 2.0  		                            &4.0		                     &2.0               & 2.0			                 \\
                Photon selection	          &2.0		                   & 2.0			                        &- 		                         &-                 &2.0			                 \\
                $\chi^2_{\rm 1C}$                         &0.7                                & -                                         &-                                &-                   &0.7             \\

                $K^0_S$ reconstruction & -     &-     &-      &1.5        &1.5      \\
                $\rho^+/\rho^0$ requirement      &-		                   & -   		                            &2.8		                     &-                 &5.1			                 \\
                $U_{\rm miss}$ requirement 	          & 0.8		                   &1.5	                            &0.9		                     &1.0               &1.0			                 \\
                Model                         &0.5                         & 0.6                                    &0.8                             &0.6               &1.0                           \\
                Branching fraction 		              &0.8		                   & 0.9 		                            &0.8		                     & 1.5              &1.5		                     \\
                $N_{J/\psi}$  &0.5		                   & 0.5  		                            &0.5		                     &0.5               &0.5			                 \\
                MC statistics           &0.3                    &0.3                            &0.3                 &0.3           &0.5                                                            \\     \hline
                Total  			      &3.8		                   & 4.0	                            &6.8		                     &5.1             &7.6 	                     \\\hline\hline
            \end{tabular}}
    \end{table*}

    The sources of systematic uncertainties on the branching fraction measurements of $\jpsi\to\bar{D}^0\pi^0$, $\jpsi\to\bar{D}^0\eta$, $\jpsi\to\bar{D}^0\rho^0$, $\jpsi\to D^-\pi^+$, and $\jpsi\to D^-\rho^+$ are classified into two types: additive and multiplicative. Multiplicative ones are involved in efficiency determination, and they are summarized in Table~\ref{sys:tot}; additive ones affect the signal yield determination, such as background shapes and signal shapes in signal yield fits. 

    The uncertainties due to tracking and PID efficiencies for kaons and pions are determined by analyzing doubly-tagged $D^{+}D^{-}$ hadronic events from $\psi(3770)$ data~\cite{bes3:kpi2017}. Using partially reconstructed hadronic decays of $D^{+}\to K^{-}\pi^{+}\pi^{+}$ and $D^{-}\to K^{+}\pi^{-}\pi^{-}$ where one $\pi^{-}$ or $K^{+}$ meson is not reconstructed, the uncertainties are estimated to be 1.0\% per track. 
    In addition, the uncertainty from the electron tracking efficiency is studied using a control sample of radiative Bhabha events $\ee\to\gamma\ee$ produced at $\sqrt{s}=3.08$~GeV, while the PID uncertainty is studied using a mixed control sample of $\ee\to\gamma\ee$ events and $\jpsi\to\ee(\gamma_{\rm FSR})$ events produced at $\sqrt{s}=3.097$~GeV.
    We quote 1.0\% and 1.0\% as the systematic uncertainties on the tracking and the PID efficiency for the electron, respectively. The uncertainty from photon detection efficiency is 1.0\% per photon, which is determined from the decays $\jpsi \to \rho^0\pi^0$ and the study of photon conversion via $e^+e^-\to\gamma\gamma$~\cite{Ablikim:2010zn}. The uncertainties of one-constraint (1C) of $\pi^0$ and $\eta$ kinematic fit are determined to be 0.7\% and 0.08\% by using the control samples $\jpsi \to p \bar{p} \pi^{0}$ and $\jpsi \to \phi \eta$, where the latter is less than 0.1\% and is negligible. 
    The systematic uncertainty associated with $K^0_S$ reconstruction is studied with control samples of the decays $\jpsi \to K^{\ast\pm}K^{\mp}$ and $\jpsi \to \phi K^0_S K^{\pm}\pi^{\mp}$~\cite{BESIII:2018jjm}. The systematic uncertainty for each $K^0_S$ is assigned as 1.5\%.
    Using the control samples $\jpsi \to \rho^+\pi^-$ and $\jpsi \to \rho^0\pi^0$, the differences in efficiencies between data and MC simulation, 5.1\% and 2.8\%, are assigned as the systematic uncertainties on mass windows of $\rho^+$ and $\rho^0$, respectively. 
    The systematic uncertainties associated with the $U_{\rm miss}$ requirement for $\jpsi \to \bar{D}^0 \pi^0, \bar{D}^0\eta, \bar{D}^0\rho^0, D^-\pi^+$ and $D^-\rho^+$ are estimated by changing the $U_{\rm miss}$ selection region from $(-0.083, 0.119)$ to $(-0.093, 0.129)$, from $(-0.050, 0.060)$ to $(-0.056, 0.066)$, from 
    $(-0.040, 0.050)$ to $(-0.044, 0.054)$, from $(-0.037, 0.040)$ to $(-0.041, 0.044)$ and from $(-0.058, 0.074)$ to $(-0.065, 0.081)$ GeV, respectively. The differences in the upper limits are taken as the corresponding systematic uncertainties.
    To estimate the systematic uncertainty due to the signal MC model, we use the ``VSS" and ``VVS\_PWAVE" models from EvtGen~\cite{Ping:2008zz} to simulate signal MC events, and the efficiency differences of the “VSS(VVS PWAVE)” and QCDF model~\cite{Sun:2015nra}, 
    assigning uncertainties 0.5\%, 0.6\%, 0.8\%, 0.6\%, and 1.0\% for $\jpsi\to D^0\pi^0, D^0\eta, D^0\rho^0, D^-\pi^+$, and $D^-\rho^+$, respectively.  The systematic uncertainties associated with the branching fractions of intermediate decays are quoted from PDG~\cite{ParticleDataGroup:2022pth}. We quote a relative uncertainty of 0.5\% determined using $\jpsi$ inclusive hadronic decays for the $N_{\jpsi}$ as the systematic uncertainty from Ref.~\cite{BESIII:2021cxx}.
    Finally, the uncertainty from the MC statistics is taken into account. The total multiplicative systematic uncertainty is determined by adding the above systematic uncertainties in quadrature. The additive systematic uncertainty due to the signal shape is negligible because it results mainly from the model as discussed earlier. The additive systematic uncertainty due to the background shape is estimated by altering the function from the first-order to the second-order polynomial, and is found to be negligible.


    \section{Summary}
    We report the first search for the weak decays of $\jpsi\to\bar{D}^0\pi^0$, $\jpsi\to\bar{D}^0\eta$, $\jpsi\to\bar{D}^0\rho^0$, and $\jpsi\to D^-\rho^+$ using $(10087\pm44)\times10^6$ $\jpsi$ events collected with the BESIII detector. With this data sample, we search for $\jpsi\to D^-\pi^+$. No evidence for any of these decays has been found. The upper limits at the 90\% C.L. on the branching fractions are determined to be: $\mathcal{B}(\jpsi\to\bar{D}^0\pi^0 + c.c.)<4.7 \times 10^{-7}$, $\mathcal{B}(\jpsi\to\bar{D}^0\eta + c.c.)< 6.8 \times 10^{-7}$, $\mathcal{B}(\jpsi\to\bar{D}^0\rho^0 + c.c.)<5.2 \times 10^{-7}$, $\mathcal{B}(\jpsi\to D^-\pi^+ + c.c.)< 7.0 \times 10^{-8}$, and $\mathcal{B}(\jpsi\to D^-\rho^+ +c.c.)< 6.0 \times 10^{-7}$. The upper limit on the branching fraction of $\jpsi\to D^-\pi^+ + c.c.$ has been improved by three orders of magnitude compared to the previous result~\cite{Ablikim:2007ag}. All results are in agreement with the SM, but more data will be helpful to test the branching fractions of these weak decays of $\jpsi$ to the order of $10^{-8}$ to constrain the parameter spaces of several theories beyond the SM.\\


    \section*{acknowledgements}

    The BESIII Collaboration thanks the staff of BEPCII and the IHEP computing center for their strong support. This work is supported in part by National Key R\&D Program of China under Contracts Nos. 2020YFA0406300, 2020YFA0406400; National Natural Science Foundation of China (NSFC) under Contracts Nos. 11635010, 11735014, 11835012, 11935015, 11935016, 11935018, 11961141012, 12022510, 12025502, 12035009, 12035013, 12061131003, 12192260, 12192261, 12192262, 12192263, 12192264, 12192265, 12221005, 12225509, 12235017; the Chinese Academy of Sciences (CAS) Large-Scale Scientific Facility Program; the CAS Center for Excellence in Particle Physics (CCEPP); Joint Large-Scale Scientific Facility Funds of the NSFC and CAS under Contract No. U1832207; CAS Key Research Program of Frontier Sciences under Contracts Nos. QYZDJ-SSW-SLH003, QYZDJ-SSW-SLH040; 100 Talents Program of CAS; The Institute of Nuclear and Particle Physics (INPAC) and Shanghai Key Laboratory for Particle Physics and Cosmology; ERC under Contract No. 758462; European Union's Horizon 2020 research and innovation programme under Marie Sklodowska-Curie grant agreement under Contract No. 894790; German Research Foundation DFG under Contracts Nos. 443159800, 455635585, Collaborative Research Center CRC 1044, FOR5327, GRK 2149; Istituto Nazionale di Fisica Nucleare, Italy; Ministry of Development of Turkey under Contract No. DPT2006K-120470; National Research Foundation of Korea under Contract No. NRF-2022R1A2C1092335; National Science and Technology fund of Mongolia; National Science Research and Innovation Fund (NSRF) via the Program Management Unit for Human Resources \& Institutional Development, Research and Innovation of Thailand under Contract No. B16F640076; Polish National Science Centre under Contract No. 2019/35/O/ST2/02907; The Swedish Research Council; U. S. Department of Energy under Contract No. DE-FG02-05ER41374.



\begin{thebibliography}{9}

        \bibitem{ParticleDataGroup:2022pth}
            R.~L.~Workman \textit{et al.} (Particle Data Group),
            \href{https://doi.org/10.1093/ptep/ptac097}{Prog. Theor. Exp. Phys. \textbf{2022}, 083C01 (2022)}.

        \bibitem{Ablikim:2006qt}
            M.~Ablikim \textit{et al.} (BES Collaboration),
            \href{https://doi.org/10.1016/j.physletb.2006.06.045}{Phys. Lett. B \textbf{639}, 418 (2006)}.

        \bibitem{Ablikim:2007ag}
            M.~Ablikim {\it et al.} (BES Collaboration),
            \href{https://doi.org/10.1016/j.physletb.2008.04.028}{Phys.\ Lett.\ B {\bf 663}, 297 (2008)}.

        \bibitem{Ablikim:2014dsn}
            M.~Ablikim \textit{et al.} (BESIII Collaboration),
            \href{https://doi.org/10.1103/PhysRevD.89.071101}{Phys. Rev. D \textbf{89}, 071101 (2014)}.

        \bibitem{Ablikim:2014fpb}
            M.~Ablikim \textit{et al.} (BESIII Collaboration),
            \href{https://doi.org/10.1103/PhysRevD.90.112014}{Phys. Rev. D \textbf{90}, 112014 (2014)}

        \bibitem{Ablikim:2017nid}
            M.~Ablikim \textit{et al.} (BESIII Collaboration),
            \href{https://doi.org/10.1103/PhysRevD.96.111101}{Phys. Rev. D \textbf{96}, 111101 (2017)}.

        \bibitem{BESIII:2021mnd}
            M.~Ablikim \textit{et al.} (BESIII Collaboration),
            \href{https://doi.org/10.1007/JHEP06(2021)157}{JHEP \textbf{06}, 157 (2021)}.

        \bibitem{Chen:2021fcb}
            S.~Chen and S.~L.~Olsen,
            \href{https://doi.org/10.1093/nsr/nwab189}{Natl. Sci. Rev. \textbf{8}, no.11, nwab189 (2021)}.

        \bibitem{Verma:1990nk}
            R.~C.~Verma, A.~N.~Kamal and A.~Czarnecki,
            \href{https://doi.org/10.1016/0370-2693(90)90507-3}{Phys.\ Lett.\ B {\bf 252}, 690 (1990)}.

        \bibitem{Hill:1994hp}
            C.~T.~Hill,
            \href{https://doi.org/10.1016/0370-2693(94)01660-5}{Phys. Lett. B \textbf{345}, 483 (1995)}.

        \bibitem{Aulakh:1982yn}
            C.~S.~Aulakh and R.~N.~Mohapatra,
            \href{https://doi.org/10.1016/0370-2693(82)90262-3}{Phys. Lett. B \textbf{119}, 136-140 (1982)}.

        \bibitem{Glashow:1976nt}
            S.~L.~Glashow and S.~Weinberg,
            \href{https://doi.org/10.1103/PhysRevD.15.1958}{Phys. Rev. D \textbf{15}, 1958 (1977)}.

        \bibitem{Datta:1998yq}
            A.~Datta, P.~J.~O'Donnell, S.~Pakvasa and X.~Zhang,
            \href{https://doi.org/10.1103/PhysRevD.60.014011}{Phys. Rev. D \textbf{60}, 014011 (1999)}.

        \bibitem{Sharma:1998gc}
            K.~K.~Sharma and R.~C.~Verma,
            \href{https://doi.org/10.1142/S0217751X99000464}{Int.\ J.\ Mod.\ Phys.\ A {\bf 14}, 937 (1999)}.

        \bibitem{BESIII:2021cxx}
            M.~Ablikim \textit{et al.} (BESIII Collaboration),
            \href{https://doi.org/10.1088/1674-1137/ac5c2e}{Chin. Phys. C \textbf{46}, 074001 (2022)}.

        \bibitem{Ablikim:2009aa}
            M.~Ablikim {\it et al.} (BESIII Collaboration),
            \href{https://doi.org/10.1016/j.nima.2009.12.050}{Nucl.\ Instrum.\ Meth.\ A {\bf 614},345 (2010)}.

        \bibitem{Yu:IPAC2016-TUYA01}
            C.~H.~Yu {\it et al.},
            \href{http://accelconf.web.cern.ch/AccelConf/ipac2016/papers/tuya01.pdf}{Proc.\ IPAC2016,\ Busan,\ Korea,\ May 2016,\ paper TUYA01,\ pp.\ 1014--1018}.

        \bibitem{BESIII:2020nme}
            M.~Ablikim \textit{et al.} (BESIII Collaboration),
            \href{https://doi.org/10.1088/1674-1137/44/4/040001}{Chin. Phys. C \textbf{44}, 040001 (2020)}.

        \bibitem{etof}
            X.~Li {\it et al.}, 
            \href{https://doi.org/10.1007/s41605-017-0014-2}{Radiat. Detect. Technol. Methods {\bf 1}, 13 (2017)};
            Y.~X.~Guo {\it et al.}, 
            \href{https://doi.org/10.1007/s41605-017-0012-4}{Radiat. Detect. Technol. Methods {\bf 1}, 15 (2017)}.  

        \bibitem{geant4}
            S.~Agostinelli {\it et al.} (GEANT4 Collaboration),
            \href{https://doi.org/10.1016/S0168-9002(03)01368-8}{Nucl.\ Instrum.\ Meth.\ A {\bf 506}, 250 (2003)}.

        \bibitem{detvis}
            K.~X.~Huang, {\it et al.},
            \href{https://doi.org/10.1007/s41365-022-01133-8}{Nucl.\ Sci.\ Tech. {\bf 33}, 142 (2022)}.

        \bibitem{ref:kkmc}
            S.~Jadach, B.~F.~L.~Ward and Z.~Was,
            \href{https://doi.org/10.1103/PhysRevD.63.113009}{Phys.\ Rev.\ D {\bf 63}, 113009 (2001)};
            \href{https://doi.org/10.1016/S0010-4655(00)00048-5}{Comput.\ Phys.\ Commun.\  {\bf 130}, 260 (2000)}.  

        \bibitem{Sun:2015nra} 
            J.~Sun, L.~Chen, Q.~Chang, J.~Huang and Y.~Yang,
            \href{https://doi.org/10.1142/S0217751X15500943}{Int.\ J.\ Mod.\ Phys.\ A {\bf 30}, 1550094 (2015)}.

        \bibitem{Ping:2008zz}
            R.~G.~Ping,
            \href{https://doi.org/10.1088/1674-1137/32/8/001}{Chin. Phys. C \textbf{32}, 599 (2008)}.

        \bibitem{Lange:2001uf}
            D.~Lange,
            \href{https://doi.org/10.1016/S0168-9002(01)00089-4}{Nucl. Instrum. Meth. A \textbf{462}, 152 (2001)}.

        \bibitem{Chen:2000tv}
            J.~C.~Chen, G.~S.~Huang, X.~R.~Qi, D.~H.~Zhang and Y.~S.~Zhu,
            \href{https://doi.org/10.1103/PhysRevD.62.034003}{Phys. Rev. D \textbf{62}, 034003 (2000)}.

        \bibitem{RichterWas:1992qb}
            E.~Richter-Was,
            \href{https://doi.org/10.1016/0370-2693(93)90062-M}{Phys. Lett. B \textbf{303}, 163 (1993)}.

        \bibitem{bernardo:2000}
            J.~M.~Bernardo and A.~F.~M.~Smith,
            Bayesian Theory, (Wiley, 2000).

        \bibitem{uplimit}
            K.~Stenson, 
            \href{http://arxiv.org/abs/physics/0605236}{{\tt arXiv:physics/0605236}}.

        \bibitem{bes3:kpi2017}
            M.~Ablikim {\it et al.}  (BESIII Collaboration),
            \href{https://doi.org/10.1103/PhysRevD.97.072004}{Phys.\ Rev.\ D {\bf 97}, 072004 (2018)}.

        \bibitem{Ablikim:2010zn}
            M.~Ablikim {\it et al.} (BESIII Collaboration),
            \href{https://doi.org/10.1103/PhysRevD.81.052005}{Phys.\ Rev.\ D {\bf 81}, 052005 (2010)}.

        \bibitem{BESIII:2018jjm}
            M.~Ablikim \textit{et al.} (BESIII Collaboration),
            \href{https://doi.org/10.1103/PhysRevD.99.011103}{Phys. Rev. D \textbf{99}, 011103 (2019)}


    \end{thebibliography}
\end{document}